\documentclass[prb,twocolumn,aps,showpacs,superscriptaddress]{revtex4-1}
\usepackage{hyperref}
\hypersetup{
	colorlinks=true,
	citecolor=blue,
	linkcolor=blue,
	urlcolor=blue}
\usepackage{graphicx}
\usepackage{amsmath}
\usepackage{physics}
\usepackage{amsfonts}
\usepackage{amssymb}
\usepackage{bm}
\usepackage{times}
\usepackage{siunitx}
\usepackage{xfrac}
\usepackage{soul}
\begin{document}
	
	\title{Non-linear dynamical response of interacting bosons to synthetic electric field}
    
    \author{Arko Roy}
	\affiliation{INO-CNR BEC Center and Dipartimento di Fisica, 
		Universit{\'a} di Trento, 38123 Trento, Italy}
    \affiliation{Max-Planck-Institut f{\"u}r Physik komplexer Systeme, 
    N{\"o}thnitzer Stra\ss e 38, 01187 Dresden, Germany}
	\author{Soumya Bera}
	\affiliation{Department  of  Physics,  Indian  Institute  of  Technology  Bombay,  Mumbai  400076,  India}
	\author{Kush Saha}
	\affiliation{School of Physical Sciences, National Institute of Science Education and Research, Jatni, Odisha 752050, India}
	\affiliation{ Homi Bhabha National Institute, Training School Complex, Anushakti Nagar, Mumbai 400094, India}
	


\begin{abstract}
We theoretically study the non-linear response of interacting neutral bosonic gas in a synthetically driven 
one-dimensional optical lattice. In particular, we examine the bosonic analogue of electronic higher harmonic generation
in a strong time-dependent synthetic vector potential manifesting itself as the synthetic electric
field. We show that the vector potential can generate reasonably high harmonics in the insulating regime, while the superfluid regime exhibits only a few harmonics. In the insulating regime, the number of harmonics increases with the increase in the strength of the vector potential. This originates primarily due to the field-driven resonant and non-resonant excitations in the neutral Mott state and their recombination with the ground state. If the repulsive interaction between two atoms ($U$) is close to the strength of the gauge potential ($A_0$), the resonant quasiparticle-quasihole pairs on nearest-neighbor sites, namely dipole states are found to a play a dominant role in the generating higher harmonics. However, in the strong-field limit $A_0\gg U$, the nonresonant states where quasiparticle-quasihole pairs are not on nearest-neighbor sites give rise to higher harmonics.
\end{abstract}



\maketitle

\section{Introduction}\label{Introduction}

The interplay between intense laser field and matter continues to be a field of extensive research both theoretically and experimentally as it allows to decode microscopic mechanism of several physical systems such as photonic, gaseous, solid-state and quantum spin systems\cite{Vampa_2017,Shintaro_2019,lewenstein_94,ikeda_19}. The non-perturbative nature of the matter-light interaction makes the field even more promising due to its potential for exhibiting unprecedented and rich physics. For example, the generation of higher harmonics in gaseous systems\cite{Ferray_1988,Huillier_1991,Jeffrey_1992, Macklin_1993,Huillier_1993}, leading to plateaus in the energy distribution of emitted light, has stimulated research for several decades and has now 
become a key candidate for attosecond science\cite{Baltuska_2003,Krausz_2009}. Recently, the higher harmonic generation (HHG) 
has been experimentally observed and theoretically studied in solid-state systems\cite{Ghimire_2010,You_2017,Hohenleutner_2015,Schubert_2014,Luu_2015,Yoshikawa736,Ndabashimiye_2016,Liu_2016,ikeda_18,ghimire_19,koki_20} and semimetals\cite{cheng_20}. While gas phase HHG is limited to complex experimental set-ups and millijoule class laser pump, the solid-state HHG turns out to be easily achievable and 
non-destructive. Thus it emerges as a potential platform for ultrafast and short wavelength coherent light sources\cite{Ghimire_2010}. 
A great volume of work suggested that the HHG in solid-state system can be used to probe the electronic properties of a wide range of materials\cite{Vampa_2015,Mengxi_2015}. It has been also shown that engineered solid-state structure\cite{,Park_2017} can be used as a possible candidate for producing stable extreme ultraviolet (XUV) waveforms\cite{Garg_2018}. 

\begin{figure}
	\includegraphics[width=1.0\linewidth]{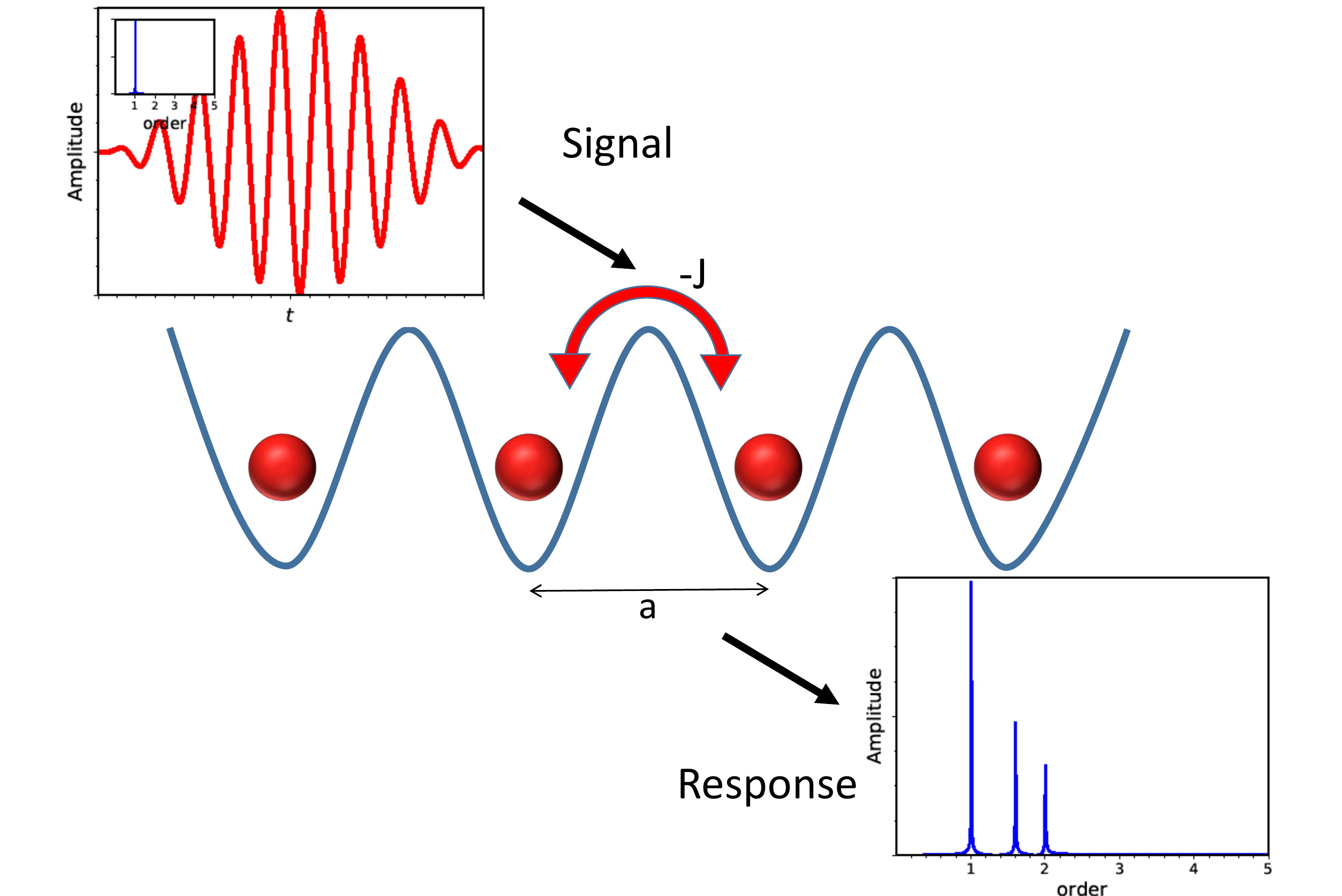}
	\caption{(Color online) Schematic diagram of light-atom interaction in an one-dimensional optical lattice setup. The synthetic 10-cycle $\sin^2$ pulse as the time-dependent gauge potential is used with a frequency $\omega/J=30$. The optical response of the system under the pulse is found to exhibit higher harmonics of the input $\omega$. Note that the coupling between neutral atom and light is synthetic in nature as also discussed in the main text.}
	\label{fig:schematic}
\end{figure}   

While most of the studies on HHG in gas and solid-state systems either assume weak interaction or those are based on single particle picture, only a few recent studies have addressed HHG in strongly correlated electronic systems such as Mott insulators \cite{Silva_2018,Murakami_2018,Nicolas_2018,Murakami_2019,Shohei_2019}. Using fermionic Hubbard model, it has been shown that the HHG can be used to resolve ultrafast many-body dynamics in a Mott insulator\cite{Silva_2018,Murakami_2018}. In the strong-field, the mechanism that leads to HHG in Mott insulators turns out to be a recombination of field-excited doublon-holon pairs with the ground states. In contrast, in the low-field limit, the itinerant doublon-holon excitations play the role in generating higher harmonics. Given HHG in interacting fermionic Mott insulators, it is natural to ask if \emph{analogous} 
higher harmonics 
can also be obtained in a bosonic Mott insulator of interacting neutral bosons loaded in optical lattice under a synthetic 
electric field. If yes, what is the underlying mechanism for the generation of such higher harmonics in the bosonic Mott insulators?
Since the typical notion of doublon-holon excitations in fermionic Mott insulators cannot be extended to bosonic systems, 
this poses a pertinent question that needs to be addressed.  Bosons in optical lattices are known to exhibit a 
class of excitations depending on the strength of the applied electric field\cite{Subir_2002}. It is also thus important to 
discern how various types of excitations contribute to HHG?
\begin{figure*}
	\includegraphics[width=0.33\linewidth]{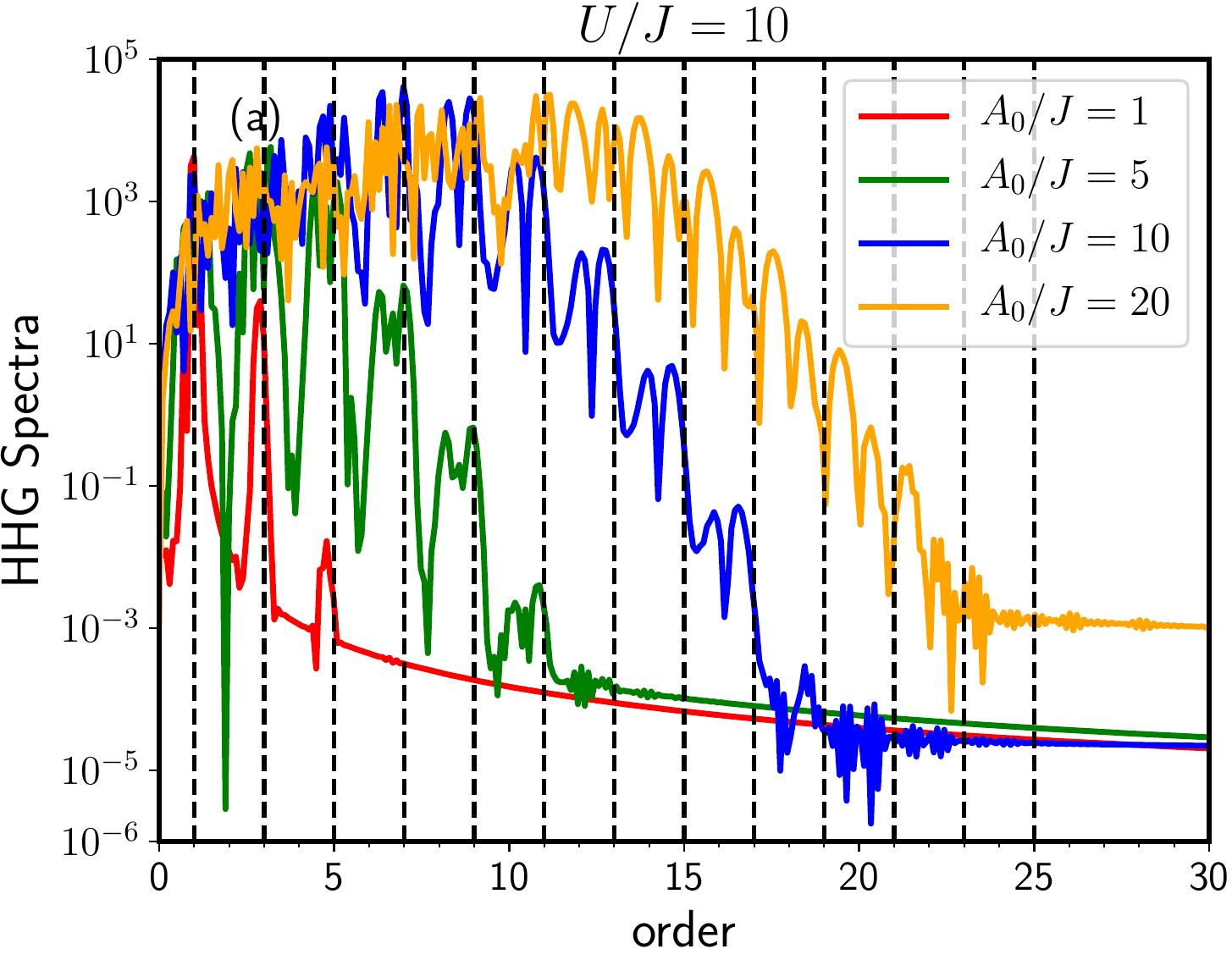}
	\includegraphics[width=0.33\linewidth]{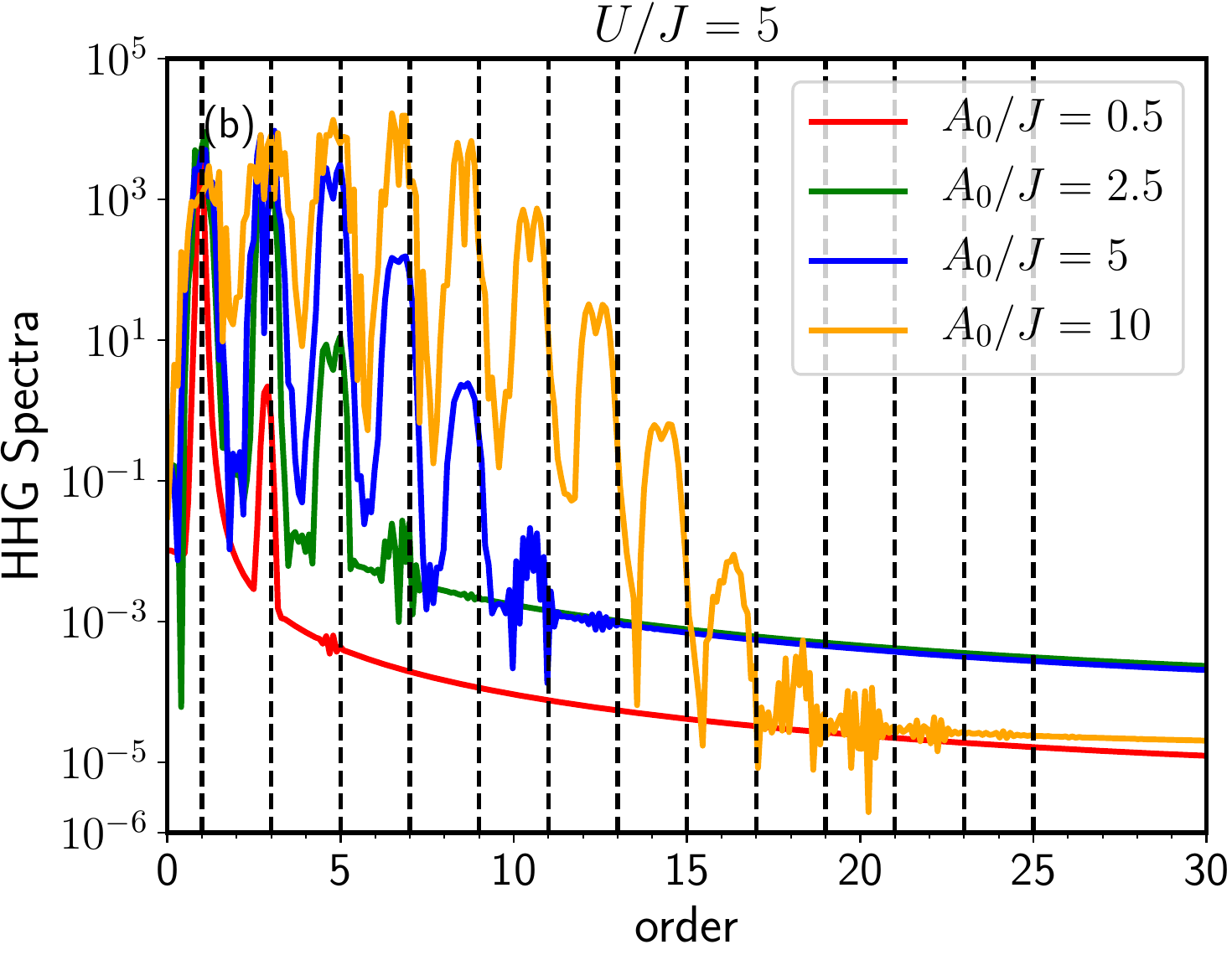}
	\includegraphics[width=0.33\linewidth]{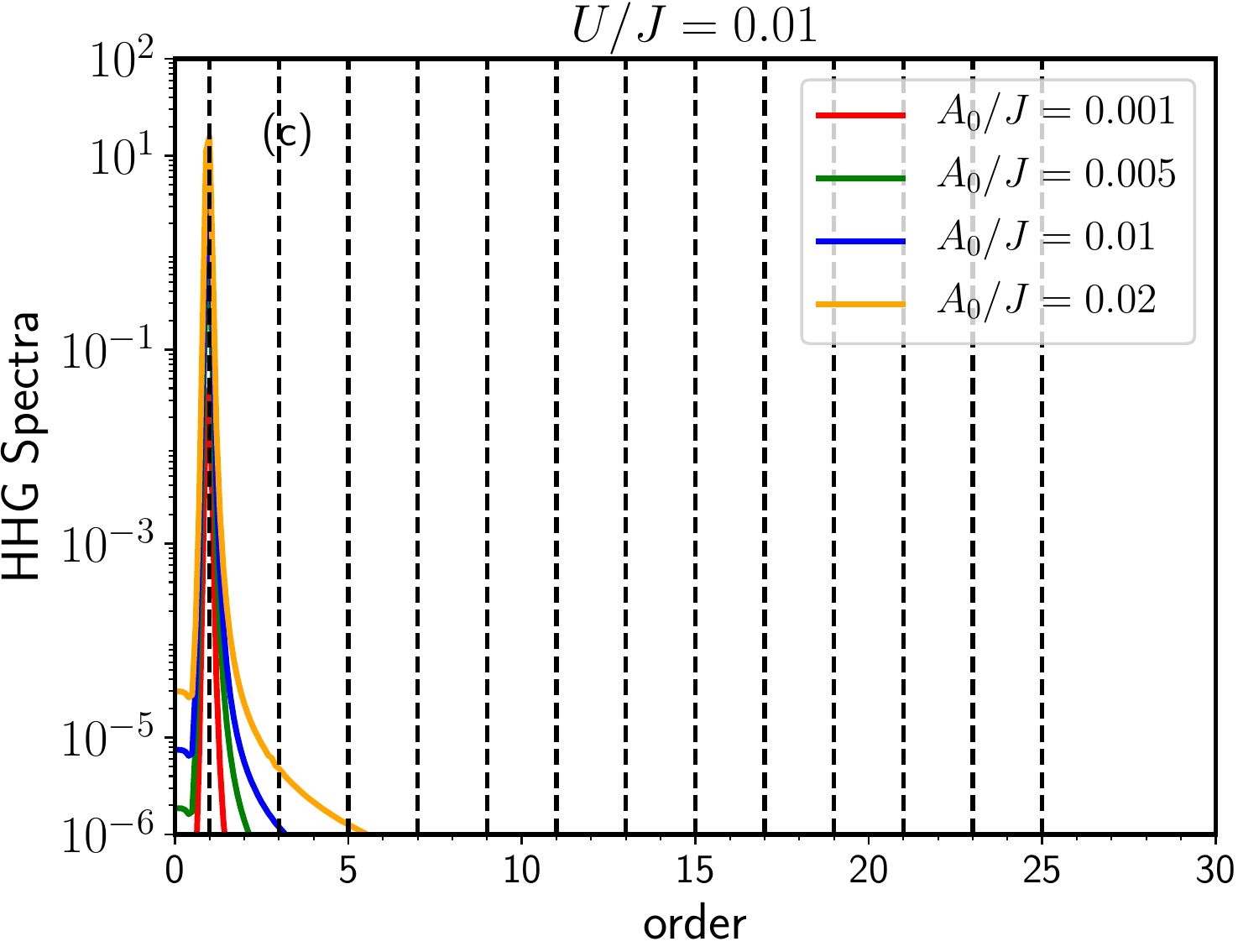}
	\caption{(Color online) Higher harmonic spectrum both in the Mott (a), intermediate (b), and superfluid (c) regime 
    for different strength of the electric field. Evidently, in the Mott and intermediate regime, the harmonic order 
        increases with the increase in electric field strength, whereas in the superfluid regime single peak appears at 
        the driving frequency. The black dashed lines indicate the odd harmonics.}
	\label{fig:hhgspectrum}
\end{figure*} 

To address the above questions, we consider one-dimensional Bose-Hubbard model and theoretically study the bosonic analogue of electronic HHG in different regimes of atom-atom interactions giving rise to the superfluid and Mott phases. We find that the dynamics of atoms in the superfluid regime is transparent to the synthetic vector potential. In contrast, 
the Mott insulating phase exhibits strikingly different optical responses. By varying the strength of the potential, 
we find the optical transitions involve an increasing order of  harmonics in the Mott insulating regime. The superfluid 
regime remains unaffected as we increase the strength of the synthetic gauge potential. We show that the generation of 
higher harmonics in the Mott insulating regime can be attributed to the formation of resonant dipole states- a pair of 
quasiparticle and keyhole on nearest neighbor sites when the potential strength is comparable with the interaction $U$. 
On the contrary, in the strong 
field limit, the HHG is attributed to the non-resonant states where  quasiparticle-quasihole pair resides on different sites.

Concerning experimental adaptability, the first challenge is to achieve a perfect
Mott insulator at finite temperature in the context of the present theoretical study. However, it has been overcome
by reducing the thermal entropy of the atoms in a novel and efficient way recently as outlined in Ref.~\onlinecite{yang2019}.
leading towards the realization of a strongly correlated Mott insulating state which is essential
for creating HHG. The particle-hole excitations which are essential for HHG, as discussed in this work, 
can be created either by the 
synthetic vector potential in the presence of additional Raman lasers, or tilting the lattice with an additional gradient 
field\cite{greiner_02}. Then the subsequent
dynamics is expected to be measured in the time-of-flight experiments as illustrated in Refs.~\onlinecite{lin_11,dalibard_11}.
We note that recently the response of the interacting ultracold atomic rubidium sample due to the application of the
femtosecond laser pulse through the resonant and non-resonant channels has been experimentally studied in 
Ref.~\onlinecite{wessels_18}. Although this cold atom set-up is driven by electrons rather than 
bosons themselves, the identification of the specific channels forms the basis of our understanding behind
the production of HHG as will be elucidated in the manuscript.

The rest of the paper is organized as follows. In Sec.~\ref{model}, we
discuss the model and formalism in the presence of a synthetic electric field. In
particular, we discuss the one-dimensional Bose-Hubbard model in the presence of a 10-cycle $\sin^2$ pulse followed 
by a discussion on the numerical methods and coupling of the electric field with the neutral atoms.  
This is followed by Sec.~\ref{results}, where we present HHG spectrum for both 
Mott and superfluid regimes for different strength of the gauge potential, and explain the possible reason for 
such HHG. We furthermore show the evolution of fidelity for different atom-atom interaction. Finally, we conclude 
with a discussion on the possible future directions and experimental implications in Sec.~\ref{conclusions}.
 

\section{Model and Methods}\label{model}
\subsection{Bose-Hubbard Hamiltonian}

We start with a gas of bosonic atoms at zero temperature trapped in an
one-dimensional (1D) optical lattice potential. The statics and dynamics of this bosonic system can well be described 
by the single (lowest) band Bose-Hubbard (BH) Hamiltonian to emulate real condensed matter systems
\begin{eqnarray}
 \hat{H} = -\sum_{\langle ll' \rangle}J\,{\hat a}_l^\dagger {\hat a}_{l'}
 +\frac{U}{2}\sum_l \hat{n}_l(\hat{n}_l -1) - \mu\sum_l n_l,
 \label{bhm}
\end{eqnarray}
where $\langle ll' \rangle$ refers to nearest-neighbors (NN) $l$ and $l'$, 
$J$ is the hopping strength between two NN sites, $U$ is the interaction, $\mu $ is the chemical potential, 
which sets the particle number $N$  in the system, ${\hat a}_l^\dagger ({\hat a}_l)$ are the 
bosonic creation (annihilation) operators with ${\hat n}_l = {\hat a}_l^\dagger {\hat a}_l$ as the occupation 
number in the $l$th lattice site. The model assumes that any excitation associated with the interaction $U$ is 
smaller than the separation energy to the first excited band of the deep optical lattice~\cite{jaksch_98,bloch_08}. 
Depending on the relative values of 
$U/J$, this model supports two distinct phases. For $U/J\gg 1 $, the system 
exhibits gapped insulating phase, namely Mott-insulator (MI) with 
commensurate integer fillings and vanishing order parameter. The Mott gap is determined by the energy difference between 
a single particle excitation band  (energy involving adding an extra particle in the Mott phase) and a single hole 
excitation band  (energy involving removing a particle in the Mott phase). Note that these particle-hole excitation 
bands are the genesis of higher harmonic orders as will be evident shortly. 
In contrast, $U/J\ll1$ leads to the  gapless superfluid (SF) phase with non-vanishing 
compressibility~\cite{fisher_89,jaksch_98,greiner_02,sansone_08}. Specifically, the presence of two gapless 
particle-hole excitation energy bands at ${\bf k}=0$ among the four constitutes the relevant superfluid 
physics~\cite{sengupta_05}.
For the present work, we consider the average particle number per site to be $1$ in the Mott phase.

\subsection{Synthetic electric field}

As opposed to the HHG in gases and solid materials where electric field, $E(t)$ naturally couples to the electrons via time-varying vector potential, $A(t)$, the {\it analogous} HHG in bosonic neutral atoms arises due to coupling between the $A(t)$ and neutral atoms in a synthetic manner. Specifically, $A(t)$ can be regulated in an experiment by detuning of two
Raman lasers which affect the mechanical momentum of the particles.
In particular, the multiple internal states of the neutral atom
are coupled with a slight off-resonant Raman lasers which in turn lead 
to dressed atoms. These dressed atoms behave like charged particles with an effective 
charge $q^{\ast}$ and they move with a finite velocity~\cite{lin_11}. This forms the basis of coupling between 
the neutral atoms and synthetic vector potential. Note that so far the experiments have demonstrated mostly
static fields. However, the gauge fields can have externally imposed time dependence giving rise to effective time-dependent
electric fields. This can be brought about periodically changing the external magnetic field, which in turn
induces time dependence on the Zeeman shift of the atomic spin states. This would change the detuning periodically. There 
have been several proposals in generating dynamical gauge fields with ultracold atoms,
however, owing to experimental limitations has hindered its progress~\cite{lin_11,goldman_14}.
Considering this experimental fact and for the sake of studying HHG, we use a $n$-cycle $\sin^2$ 
time varying vector potential of the form of a pulse, $A(t)=A_0 \sin^2(\omega\, t/2n)\sin(\omega\, t)$ 
with $\omega$ being the frequency of oscillation, where $E(t)=-\partial_t\,A(t)$.  The strength of the vector 
potential $A_0 \sin^2(\omega\, t/2n)$ smoothly 
varies with $t$ and the maximum 
value is attained at the half-cycle of the pulse. The $A(t)$ minimally couples to the system via the hopping term 
as $J\,e^{i\Phi(t)}$, where $\Phi(t)=q^{\ast}\,A(t)\, a$, with $a$ as the lattice
constant. Figure~\ref{fig:schematic} illustrates the schematic of light-atom coupling in 
1D optical lattice set-up, which is synthetic in nature.

We point out that the variation of $A(t)$ which leads to the change in momentum
with time, in turn, imparts a force on the dressed atoms and constitutes the effective particle current
as discussed in Sec.~\ref{current}.
It could be also interpreted as radiation of a neutral system. Also, we note that the $A(t)$ can couple the particle-hole excitations giving rise to various resonant and 
nonresonant excitations (located within the gap between the lowest and first excited states) 
depending upon the ratio between $U$ and $A_0$, where $A_0$ is taken in units of energy~\cite{Subir_2002}. 
This will be discussed in Sec.~\ref{excitation}.   

\subsection{Numerical methods}

To study the dynamics, we solve the time-dependent Schr\"odinger equation 
$\hat H(t)\,\psi(t)= i\hbar \frac{\partial \psi(t)}{\partial t}$ numerically.  The ground state of the interacting 
Hamiltonian (Eq.~\ref{bhm}) at $t=0$ is computed for a lattice site of length $L=7$ and total number of atoms $N=7$, 
using exact diagonalization. It is then evolved under time-dependent Hamiltonian $\hat H(t)$ using the Runge-Kutta 
algorithm choosing an optimum temporal step size which renders the dynamics convergent (see Appendix A). For a detailed 
review on the use of numerical methods to study HHG, 
the readers may refer to Ref.~\onlinecite{yu_19}. With the evolved wavefunction, various quantities are computed  
to investigate the response of the interacting many-body system to the synthetic pulse. It may be mentioned 
here that with the 
increase in the system size, the dimension of the Hilbert space increases exponentially, and thus computing
the dynamics becomes computationally expensive. We have thus refrained here from providing results for large system
sizes, although, we have checked that the mechanism for HHG generation remains unchanged with the increase in the
system size which is discussed in Appendix B.

\section{Results and discussions}\label{results}
\subsection{HHG spectrum}
\label{current}

To investigate the effect of light field, we first evaluate current operator defined as 
\begin{align}
 \mathcal{J}(t)=-i\,a\,q^{\ast}\,J\sum_{i=1}^L \left(e^{-i\Phi(t)}b_{i}^{\dagger}b_j-{\rm H.c.}\right).
\end{align}	

\begin{figure}
    \includegraphics[width=0.99\linewidth]{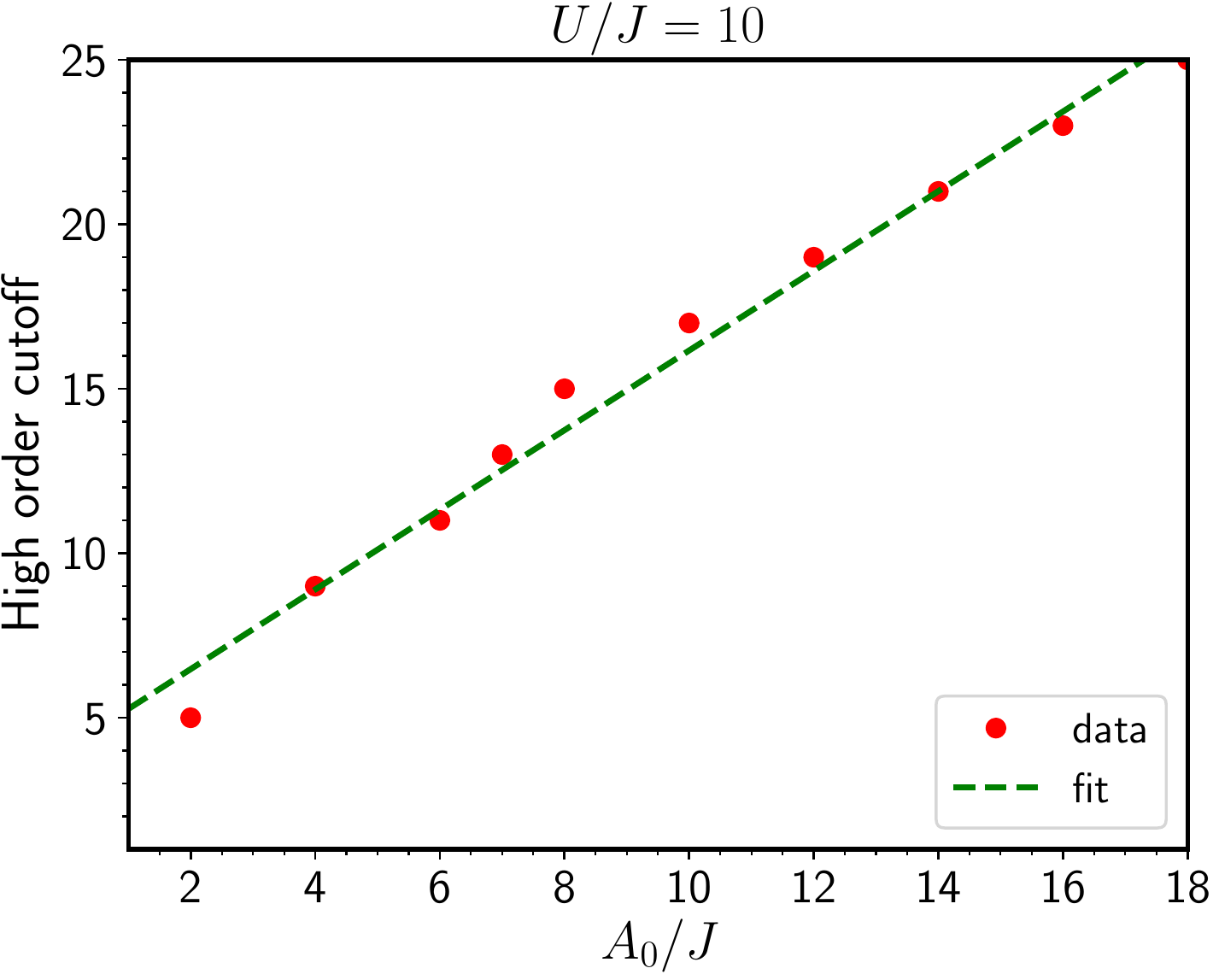}
    \caption{(Color online) Dependence of the high-order harmonic spectrum on the strength of 
              the gauge potential for $U/J =10$ revealing a linear cutoff law.}
    \label{cutoff}
\end{figure}

The HHG spectrum is obtained from the dipole acceleration $F(t)=d\mathcal{J}/dt$ in frequency space. 
Note that the non-linear current generated due to synthetic driving in this present
study does not emit photons, in contrast to the real materials.
Figure~\ref{fig:hhgspectrum} shows HHG spectra in deep Mott insulating, intermediate and deep superfluid regimes 
for different strength of the gauge potential $A_0$. Evidently, there is a single dominant peak in the superfluid 
regime and the location of the peak does not change with the change in the strength of the potential. 
This is attributed to the lowest quasienergy band 
current without any higher quasienergy curve crossings, and is usually 
typical of any single band Hamiltonian under periodic driving\cite{Krausz_2009}. 
As mentioned before, since the superfluid phase is gapless, there is only one band and the 
application of synthetic electric field leads to typical Bloch-type oscillation within this band.
In contrast, the Mott insulating phase, which is characterized by a gapped excitation spectrum, exhibits  
optical transitions between the quasienergy levels brought about by the coupling of particle-hole excitations. 
This is further associated with higher harmonics of the synthetic driving 
frequency. When $U/J=5$, that is, in the 
intermediate regime, higher harmonics are also generated with the 
application of the synthetic light pulse as evident from 
Fig.~\ref{fig:hhgspectrum}b. However, the value of the highest order 
harmonic for any $A_0/J$ is less than what is obtained in the deep insulating regime. Furthermore, it is 
to be noted here that unlike the 
fermionic system studied in Ref.~\onlinecite{Silva_2018}, as we increase the synthetic field 
strength $A_0$, the order of harmonics increases monotonically. The cutoff scaling law with the strength
of the vector potential $A_0$ turns out to be linear as shown in Fig.~\ref{cutoff}, corroborating the typical 
scaling obtained in gases and solids\cite{lewenstein_94,Ghimire_2010}.

For low order harmonics, the peaks appear exactly at odd-integer multiple, however, they deviate slightly for
high order harmonics. As we increase the strength of the gauge field, the deviation starts to increase.
This may be attributed to the unitary dynamics of the model that we consider. The absence of the bath induced dephasing
terms in the current Hamiltonian leads to the offset of the odd harmonic spectrum for the higher orders
~\cite{vampa_14,gentile_14,Silva_2018}.
Cold atoms in optical potentials provide extremely clean conditions and do not suffer from dissipation
induced by the coupling to phonons. However, dissipation can be introduced to get clear HHG peaks
through non-Hermitian Hamiltonian, which is beyond the scope of our present study.
Furthermore, in some cases, the prominent splitting of the peaks may be attributed to the current due to
the coupled quasi-energy levels
similar to the two-bands Bloch oscillation in a static field, where the dynamics of the 
two-band systems is 
characterized by two timescales as explained in 
Ref.~\onlinecite{Breid_2007}.

The reason for the HHG in the bosonic system may be associated with the recombination of resonant 
and non-resonant excitations  with the ground state in the presence of synthetic vector potential.
These excitations do not have analogues to the real-systems, and the mechanism of HHG in real solids
and gases is different. It is worth pointing out here that the energy scales
for our current set-up should not be strictly compared to that of real solid-state systems or gases. 
Whether be it cold-atom or real materials, since the related physics should be scale-independent, we use 
normalized dimensionless units in all the relevant plots and discuss the mechanism. Thus, to understand the 
role of these excitations in HHG in cold atom setup, we first briefly 
review typical excitations in the Mott phase of an optical lattice in the presence of a static electric field.

 \subsection{Excitations in the Mott insulating phase}
\label{excitation}
 
 A static electric field $E$ introduces the Stark term  $H'=E\sum_{i}i\, a_i^{\dagger}a_i$ in Eq.~(\ref{bhm}), 
 which is equivalent to  tilt or a linear potential gradient in the optical lattice in the coordinate space. 
 We also note that the addition of $H'$ is the same as modifying hopping $J$ by 
 Peierls phase involving electric field as discussed before. They are related by a simple gauge transformation 
 as shown in Ref.~\onlinecite{Krieger_1986}. It is 
 convenient to understand the excitation spectrum of the system and subsequent tunneling between the sites 
 in terms of tilting of the optical lattice. 
\begin{figure}
	\includegraphics[width=0.99\linewidth]{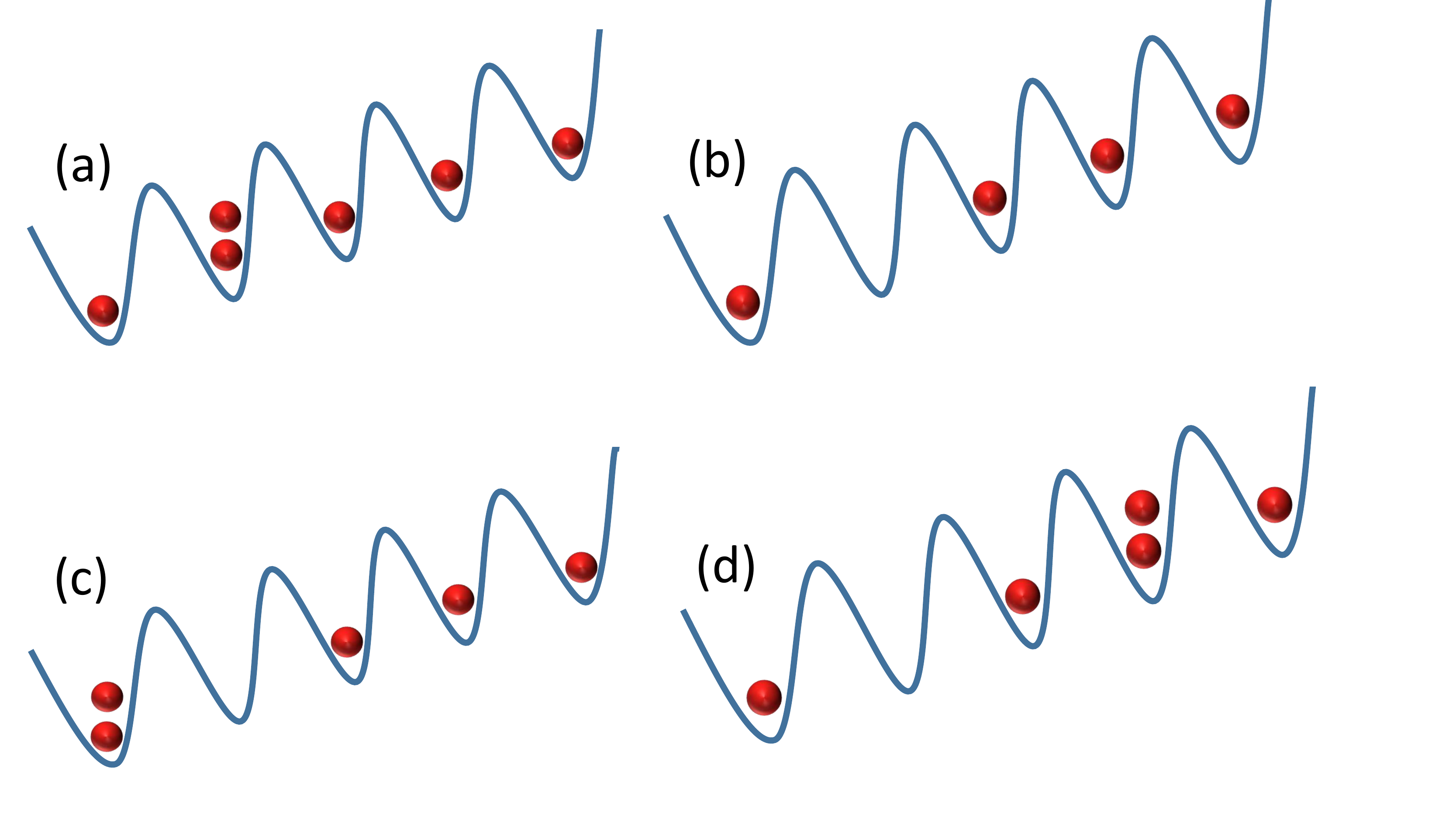}
	\caption{(Color online) Schematic representation of deformed Mott insulators in a field-driven tilted optical lattice. a) a quasiparticle state at site 2, b) a quasihole state at site 2, c) a resonant state with a pair of quasiparticle-quasihole states on nearest-neighbor sites. This state is usually called dipole state, and d) A non-resonant states with a quasiparticle-quasihole pair are not on nearest-neighbor sites.}
	\label{fig:excitations}
\end{figure}
In the Mott limit, i. e., $J\ll U$, the typical ``quasiparticle" and ``quasihole" excitations can be created by adding 
a single particle on a site or removing one particle from a site (see Fig.~(\ref{fig:excitations})), respectively. 
Such quasiparticle or quasihole states over the Mott states turn out to be localized even in the presence of any 
finite {gauge potential}\cite{Subir_2002}. Consequently, these states do not extend across the whole system to 
produce significant changes in the initial state. Thus such excitations of deformed Mott state with net finite charge 
cannot take part in generating higher harmonics. On the other hand, Mott state with zero net charge, usually called 
neutral Mott state produces various families of excitations in the presence of a synthetic electric field. The 
excitations of such Mott state are possible only if $E\sim U$, where $E$ is measured in units of energy. 
In this limit, a quasiparticle-quasihole pair is 
formed on nearest-neighbor sites and they may tunnel resonantly into the nearest site. These states are called dipole 
states\cite{Subir_2002} and differ in energy from the Mott state by $\sim E-U$ when $J=0$. Thus at $U=E$, states become 
degenerate and an infinitesimal $J$ leads to a resonant coupling between them. In addition to these, there are states 
where a quasiparticle-quasihole pair is not on nearest-neighbor sites. Such states are called non-resonant states. For 
further reading, the reader can consult Ref.~\onlinecite{Subir_2002}.

Here we provide a representative example to identify resonant subspace for a system with $L=7$ sites and $N=7$ atoms. In 
this case, the neutral Mott state has one atom in each site, i. e.,  $|1111111\rangle$. At $E\sim U$, this state is 
coupled to the single dipole states $|1201111\rangle$, $|1021111\rangle$, etc. These dipoles are then coupled to states 
with two dipole $|1201201\rangle$, $|1021021\rangle$, etc. These are further coupled to states with three dipole, 
and so on. Note that these multiple dipole states over the Mott states are part of the resonant family. As mentioned, 
the non-resonant states which are not made up of nearest dipoles can be expressed 
as $|1012111\rangle$, $|1101121\rangle$, etc,  We note that the resonant and non-resonant states 
in the present scenario 
constitute a fraction of the total number of states. With this, we now aim to find the evolution 
of different resonant and non-resonant states in the presence of a synthetic vector potential with strength $A_0$, 
playing the role of $E$.

 \begin{figure*}
     \includegraphics[width=0.33\linewidth]{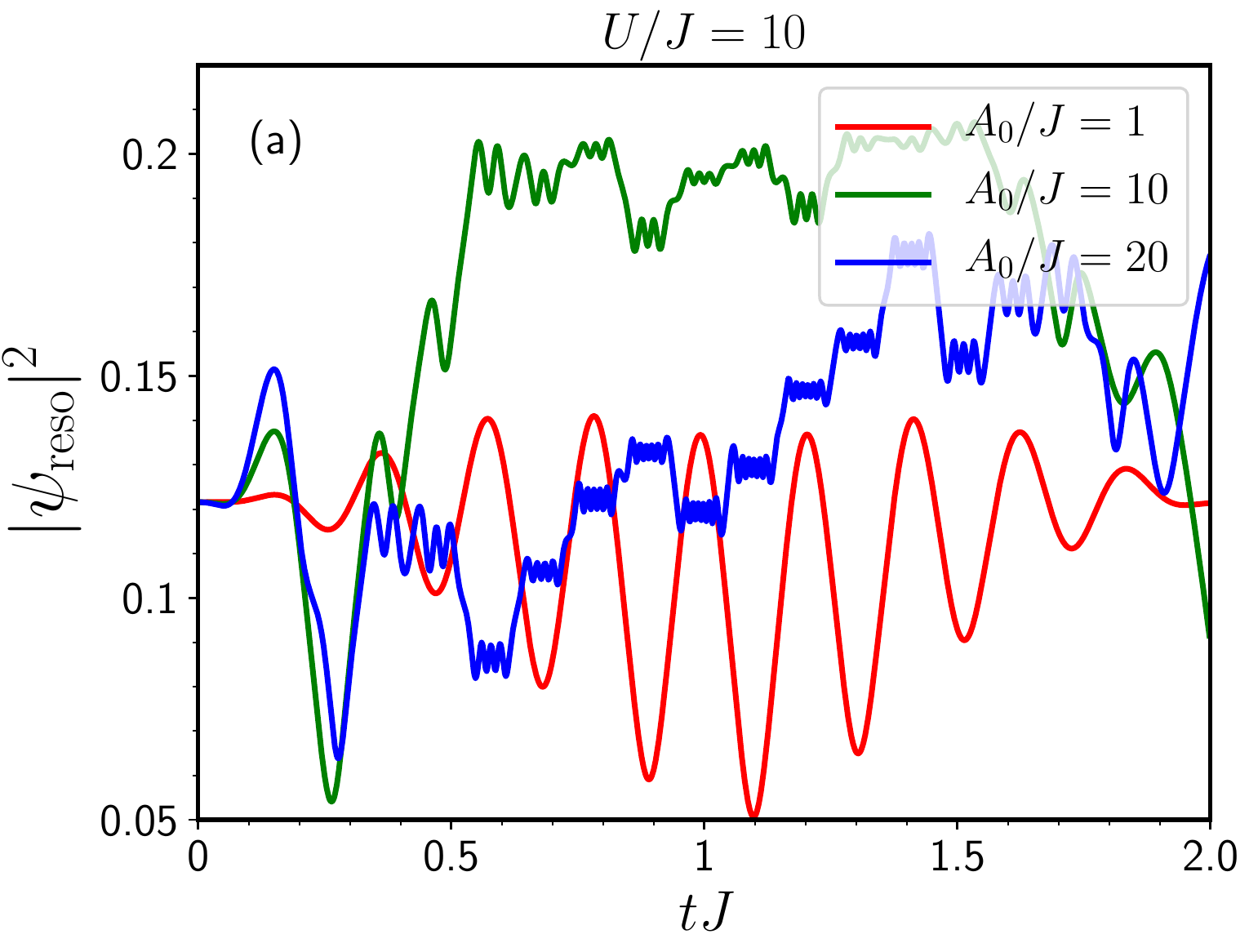}
        \includegraphics[width=0.32\linewidth]{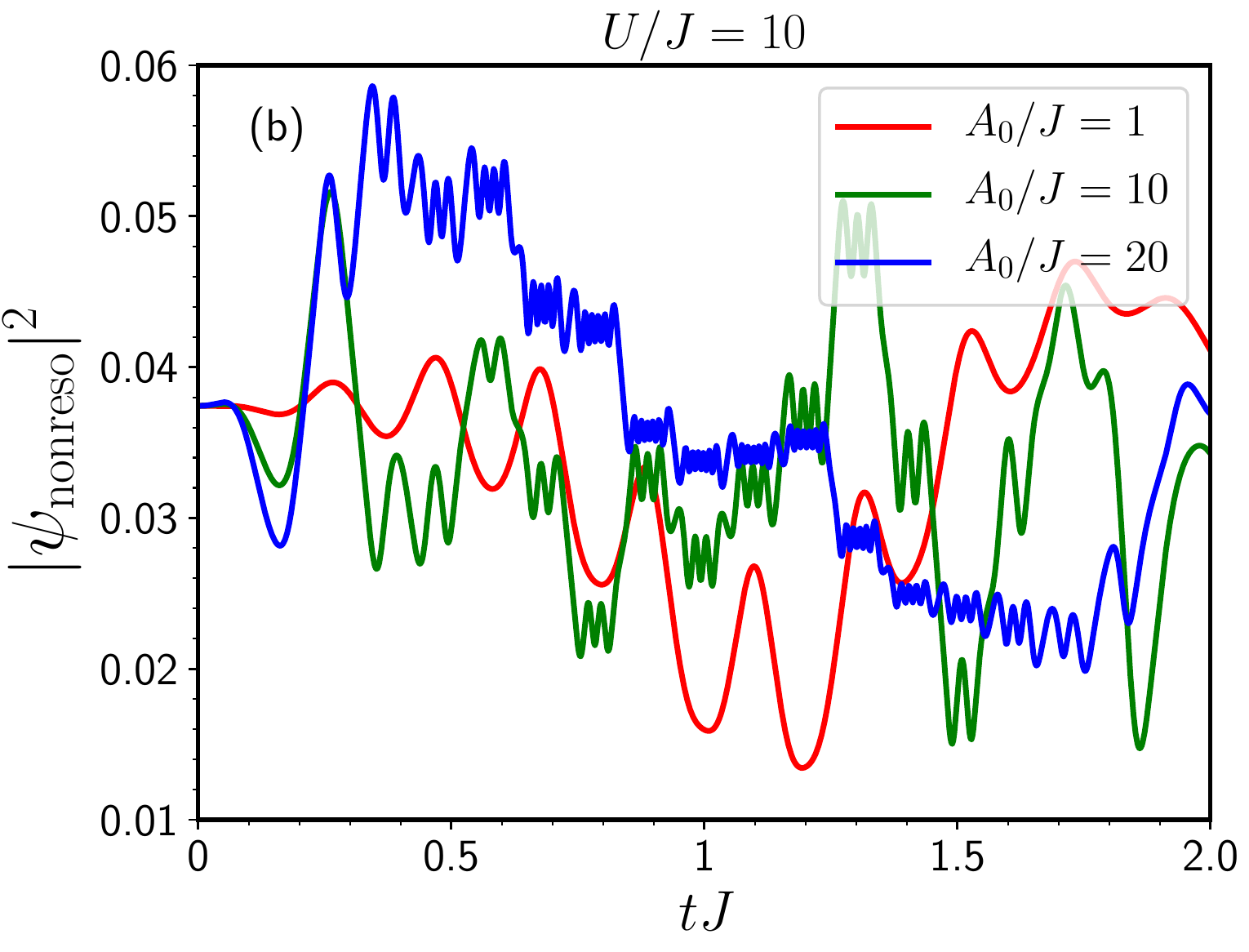}
         \includegraphics[width=0.34\linewidth]{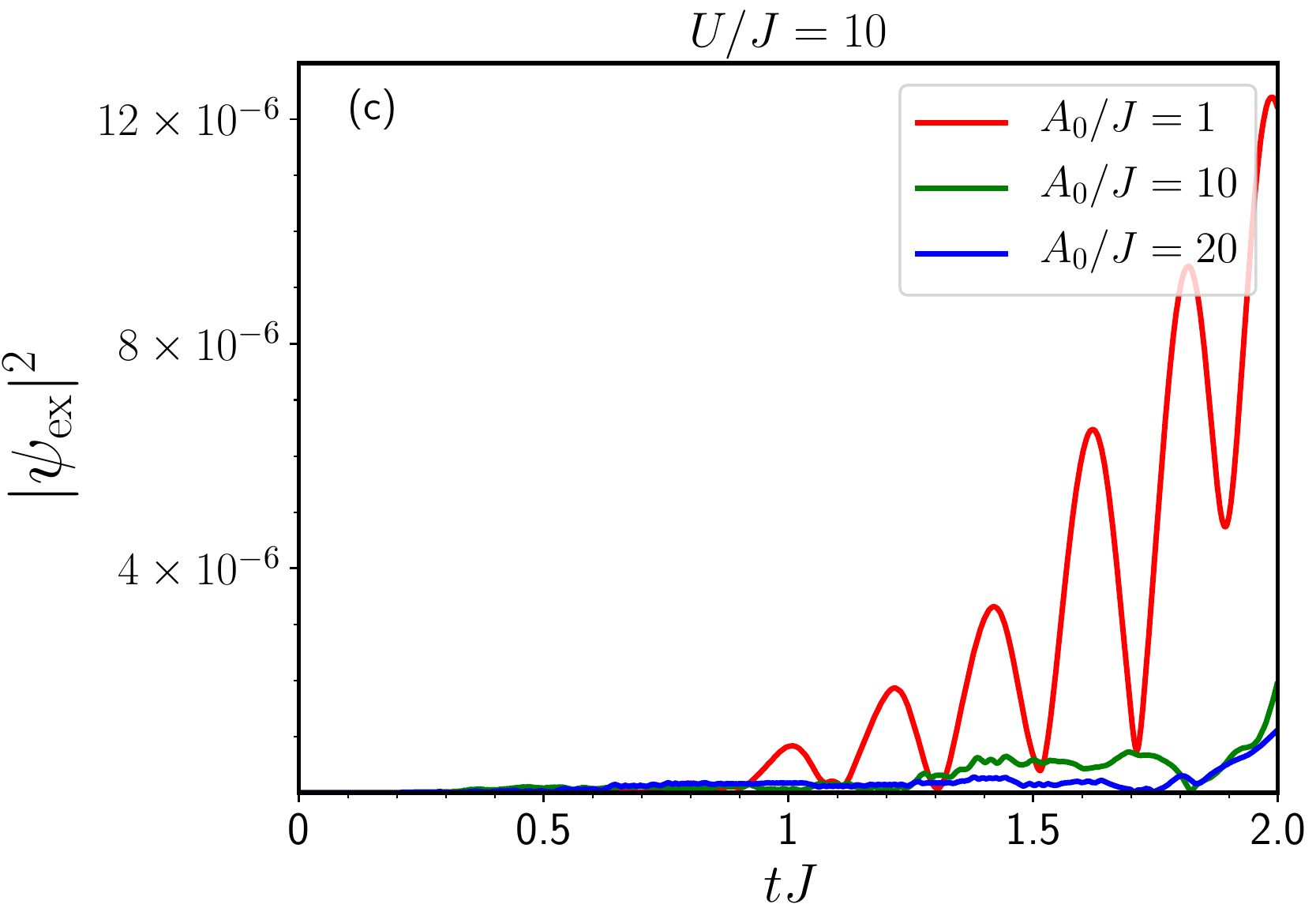}
                 \caption{(Color online)  Evolution of probability of finding (a) $|2011111 \rangle$ (single dipole state), (b) $|1012111\rangle$ (non-resonant states) and (c) $|5200000\rangle$ (highly excited state) states for varying strengths of $A_0/J$ in the Mott insulating regime with $U/J=10$. Evidently, the evolution probability  for the resonant dipole states (green curve in (a)) is found to be significant near $U/A_0\ll 1$, whereas for $U/A_0\gg 1$, the non-resonant state (blue curve in (b)) takes over the resonant one. The highly excited states are always suppressed (all curves in (c)) irrespective of the strength of light field. 
                                               }
        \label{fig:reso_nonresonant}
 \end{figure*}

\subsection{Mechanism for HHG}

Having discussed the possible excitations in the Mott phase, 
we now investigate overlapping of resonant ($|\psi_{\rm reso}|^2$)  and non-resonant states ($|\psi_{\rm nonreso}|^2$) 
with the evolved ground state $\Psi_{G}(t)$ of the system for different strength of the gauge potential. 
Indeed, this will help us quantifying excitations responsible for HHG spectra as shown in Fig.~\ref{fig:hhgspectrum}. 
For illustration, we  consider one resonant $|2011111\rangle$, one nonresonant $|1012111\rangle$ and one 
highly excited $|5200000\rangle$ state. Fig.~\ref{fig:reso_nonresonant} shows the 
evolution of the probability of finding these states due to the application of the laser pulse. 
For weak strength of the potential $A_0\ll U$, 
the $|\psi_{\rm reso}|^2$ oscillates  follows the synthetic vector potential and the magnitude is very 
small. However, for $A_0\sim U$, we see enhanced $|\psi_{\rm reso}|^2$, whereas $|\psi_{\rm nonreso}|^2$ has an order of 
magnitude less contribution to the system.
This is shown in Fig.~\ref{fig:reso_nonresonant}(a). Thus the appearance of HHG can be attributed to the 
recombination of dipole states with the 
ground state at $A_0\sim U$. In contrast, 
the $|\psi_{\rm reso}|^2$ reduces with increasing $A_0$, but the contribution for non-resonant state starts 
to dominate as 
evident from the blue line in Fig.~\ref{fig:reso_nonresonant}(b). Thus, as expected, the HHG for stronger field is no 
longer due to the 
dipole states, rather the non-resonant states start to play a vital role. Apart from the resonant and non-resonant states,
the excitation spectrum also consists of arbitrary highly excited states, for example, $|5200000\rangle$.
During the dynamics, the probability of contribution from these states is exceedingly low as is evident 
from Fig.~\ref{fig:reso_nonresonant}(c) even when $A_0\geqslant U$. 
This trend is different from the behaviour of resonant and non-resonant states in the spectrum. We find similar
qualitative behaviour for the other resonant, non-resonant, and highly excited states to the application of synthetic pulse
with different strengths (not shown).

\subsection{Loschmidt Echo}
\begin{figure}
    \includegraphics[width=0.98\linewidth]{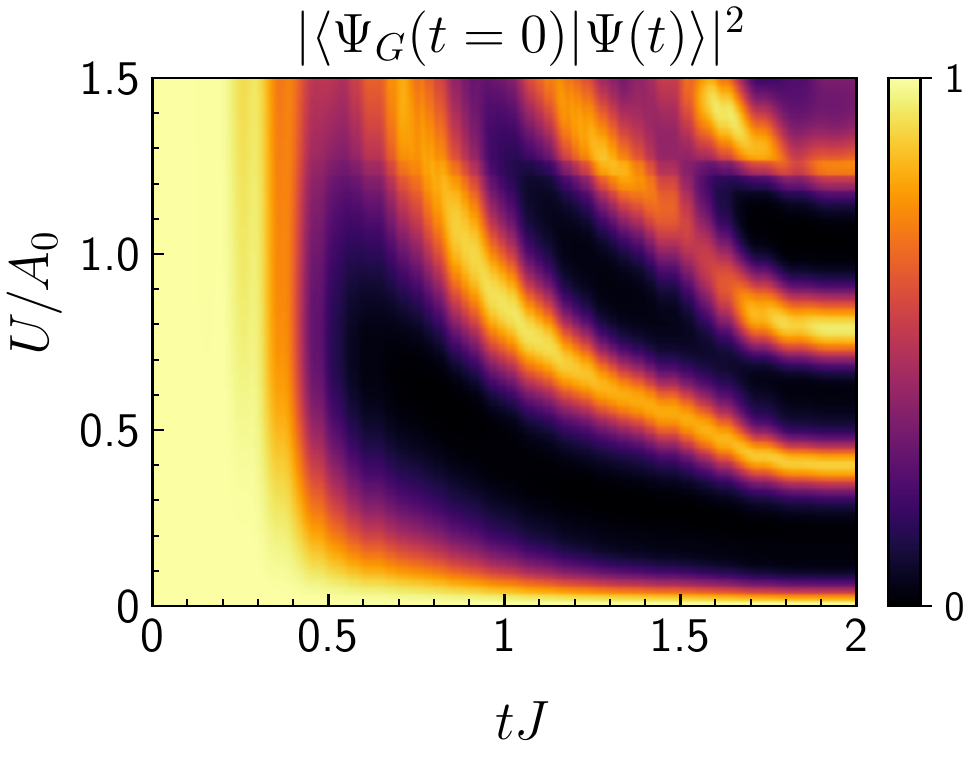}
            \caption{False color coded image showing the variation in Loschmidt echo
            $|\langle\Psi_G(t=0)|\Psi(t)\rangle|^2$ with time for a fixed value of $A_0/J$ and varying $U/A_0$. In the 
            limit of $U=0$, the initial ground state does not change appreciably, corroborating the localised nature of 
            the ground state in the presence of light field\cite{Subir_2002}. In contrast, Mott regime shows 
            reentrant behavior due to recombination of excited states with the ground state in the presence of synthetic
            vector potential. This in turn leads to higher harmonics in the emitted spectra as elaborated in the main text.
                }
    \label{fig:population}
    \end{figure}

To further analyze the dynamics towards the production of HHG, we turn our study to the evolution of ground state 
population or Loschmidt Echo i. e., $|\langle\Psi_G(t=0)|\Psi(t)\rangle|^2$ for fixed $A_0/J > 1$ and with 
varying $U/A_0$. It is evident 
from Fig.\ref{fig:population} that in the early stages of the laser pulse $tJ \sim 0.4$, the ground state remains 
unaffected, that is, the population is close to unity. When $U=0$, that is, in the superfluid regime, any
finite $A_0$ renders the single-particle wavefunction to be localized. The application of synthetic potential  
does not bring any substantial changes to the initial ground state. It is worth mentioning here that the effective 
amplitude $A_0(t)$ reaches maxima only at the half-cycle ($tJ=1$) of the pulse. 
In the domain $U/A_0 < 1$, the non-resonant states
along with the resonant dipole ones start to contribute. During the period $tJ \sim 0.4$, the potential is weak to excite
these transitions, hence the population of the evolved state does not change. For $tJ > 0.4$, the amplitude of the
vector potential becomes effectively strong to initiate modifying the population of the ground state. The 
initial state at $t=0$ gets deformed. As the vector potential becomes comparatively weak after the half-cycle,
the system tries to relax back to the ground state giving rise to a re-entrant behaviour. We now focus our attention on 
the regime
when $U \sim A_0$. The dipole formation starts around the mid-cycle, and the initial Mott state gets deformed giving rise
to a decrease in the population. Following the sinusoidal nature of the pulse, we find re-entrant behaviour of the ground
state. The excited state relaxes back and recombines with the ground state. It may be recalled here that $U \sim A_0$
is a necessary condition for the production of HHG. As $U/A_0 > 1$, the deformation of the initial
Mott state gets plagued by weak vector potential. Hence, instead of the population going to zero, it remains as a finite
quantity.

\section{conclusions}\label{conclusions}

We have performed a detailed analysis to show that the Mott phase of the BHM admits HHG production, whereas the SF phase
does not. In particular, we have demonstrated here the mechanism behind the generation of higher harmonics using intense 
synthetic electric field
in an interacting bosonic gas loaded in an one-dimensional optical lattice. We find that the strong synthetic 
time-dependent vector potential manifesting itself as the synthetic electric field can generate 
reasonably high harmonics in the insulating regime, while the superfluid regime is transparent to it. In the insulating 
regime, the order and number of harmonics increase with the variation in the strength of the synthetic vector potential 
translating to a linear cutoff law. This is attributed to 
the field-driven resonant and non-resonant excitations  in the neutral Mott state and their subsequent recombination with 
the ground state. We have shown that if the repulsive interaction between two atoms ($U$) is close to the strength of 
the potential ($A_0$), the resonant quasiparticle-quasihole pairs on nearest-neighbor sites play a pivotal role 
in generating higher harmonics. However, in the strong-field limit $A_0\gg U$, the nonresonant states where 
quasiparticle-quasihole pairs are not located on the nearest-neighbor sites contribute to higher harmonics. Finally, we 
have indicated a possible experimental scope of the obtained results. In the future, investigation of HHG generation
in square lattices would be a natural choice; where identifying and understanding dipole states in higher dimensions is 
itself a topic of fundamental importance. We believe that on one hand, the creation of a near Mott insulator and 
on the other, the experimental feasibility to study strong-field ionization in ultracold atoms~\cite{wessels_18}
would make it feasible to study analogous HHG by bosons and its variants related to strong field 
physics in the near future.
\begin{acknowledgments}
AR  acknowledges  support  from  Provincia  Autonoma  di  Trento. 
SB would like to thank G. Dixit, F. Evers, and S. Pujari for several
discussions. SB acknowledges support from DST, India, through
Ramanujan Fellowship Grant No. SB/S2/RJN-128/2016,  through Early
Career Award ECR/2018/000876, and MPG for funding through the Max
Planck Partner Group at IITB. We thank the visitor program of Max
Planck Institute for the Physics of Complex Systems, Dresden for
hospitality during the initial stages of the work. 
We also thank the anonymous referees for their thorough review and valuable 
comments,  which contributed to improving the quality of the manuscript.
\end{acknowledgments}

\appendix
\label{appendixA}
\section{Convergence check of HHG spectrum for different temporal steps of Runge-Kutta algorithm}
\begin{figure}
		\includegraphics[width=0.98\linewidth]{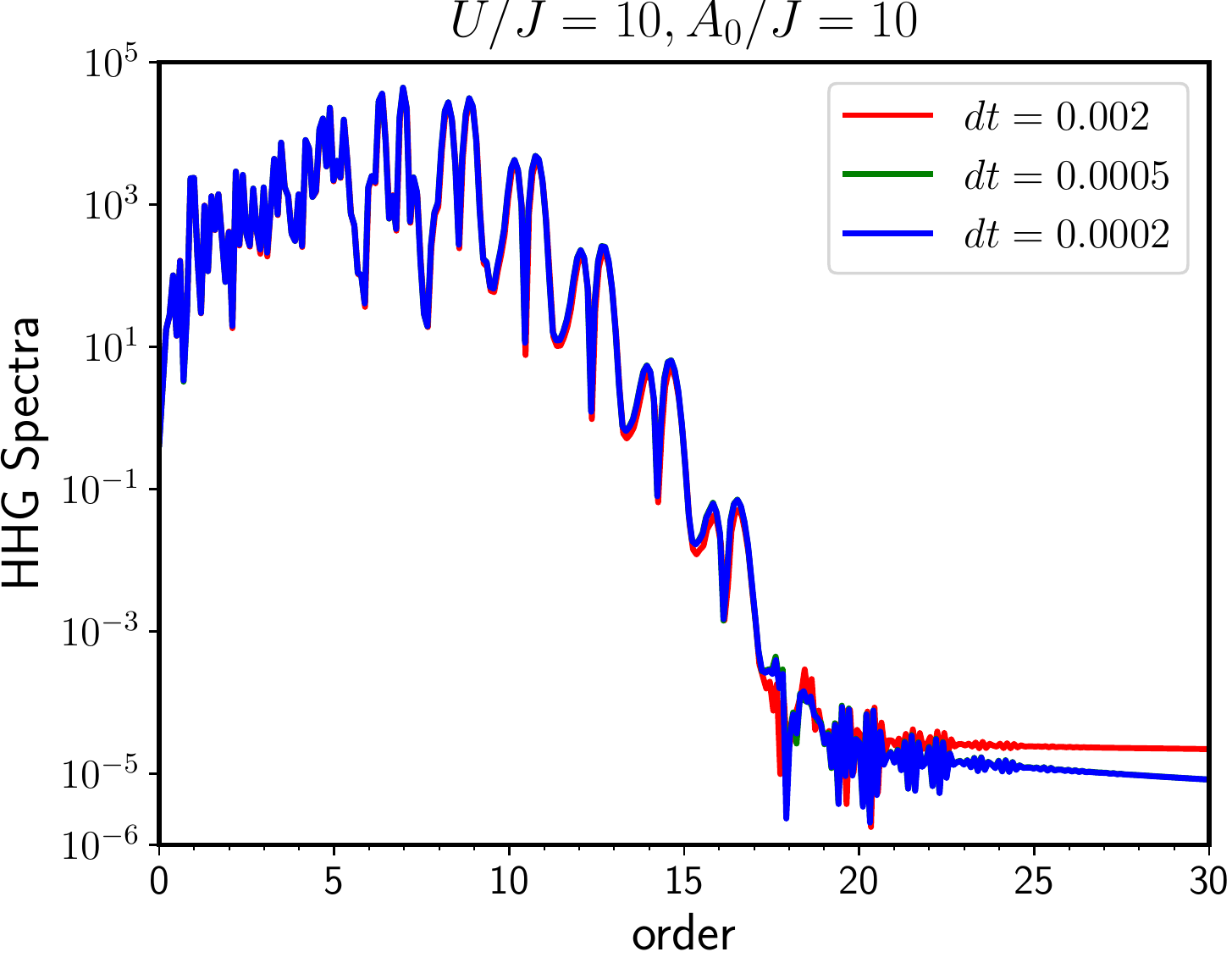}
		\caption{Plots of HHG spectrum in the deep Mott insulating regime for different temporal steps used in the 
                 Runge-Kutta algorithm. Evidently, the plots for different temporal steps match well with other, 
                 justifying the reliability, accuracy of the obtained dynamical behavior. }
		\label{fig:appnxA}
	\end{figure}

In this appendix, we provide concrete evidence of numerical accuracy and convergence of Runge-Kutta algorithm used in this present study. For different temporal steps in Runge-Kutta algorithm, the HHG spectrum in Fig. ~\ref{fig:appnxA} is found to fall on top of each other, justifying the accuracy and convergence of the dynamical behavior. 	

\label{appendixB}
\section{Dependence on system size}
\begin{figure}
        \includegraphics[width=0.98\linewidth]{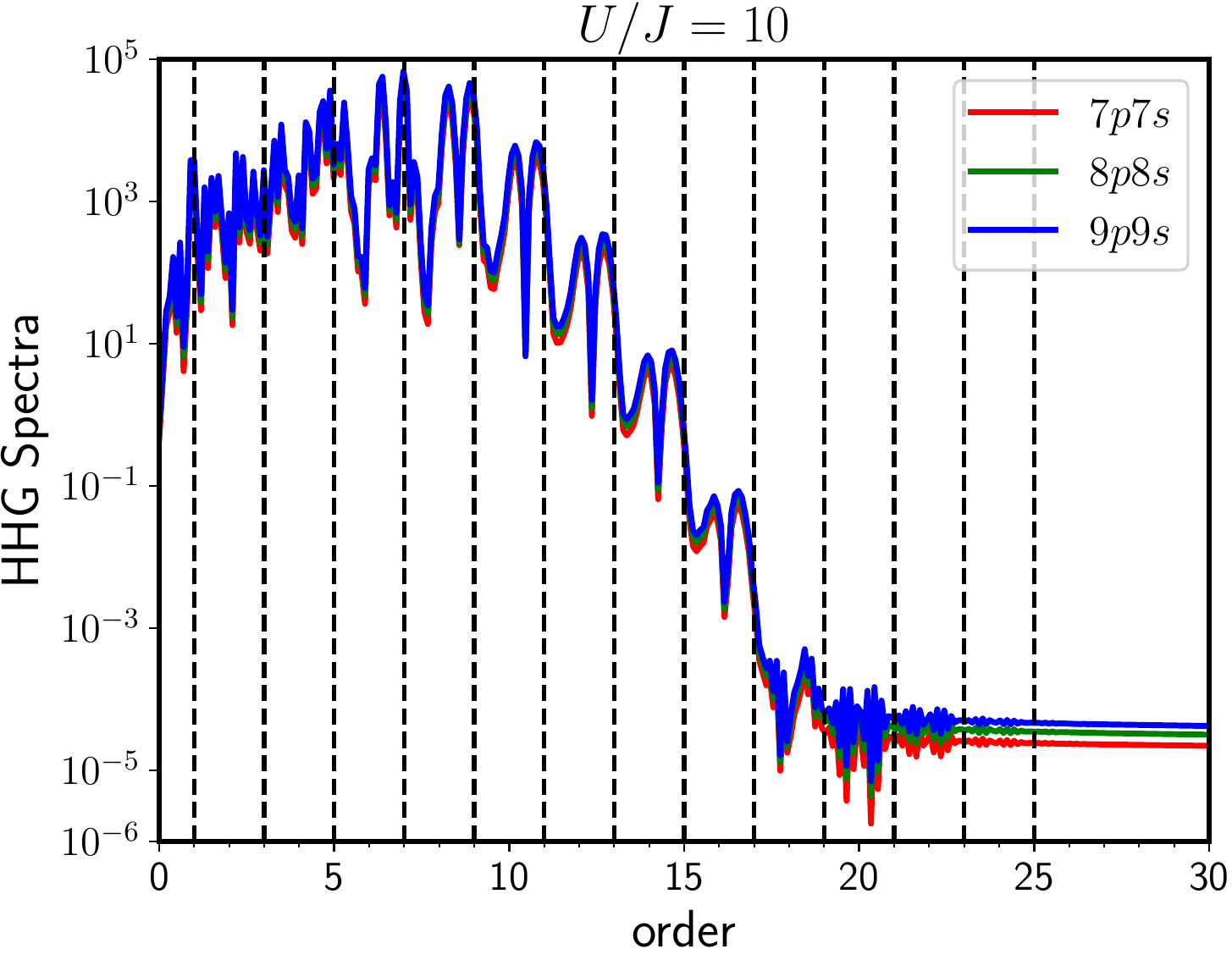}
        \caption{Plots of HHG spectrum in the deep Mott insulating regime for $A_0/J=10$ for different system sizes ranging
                 from 7 particles(p) in 7 sites(s) to 9 particles(p) in 9 sites(s). Evidently, the plots for different
                 configurations match well with other, justifying that our results are independent
                 of system sizes.}
        \label{fig:appnxB}
    \end{figure}
We provide here additional results to support that our calculations are devoid of any finite size effects for the
case $U/J=10$ and $A_0/J=10$. As mentioned before, we used ED for our computations and going beyond the configuration
with 9 particles and 9 sites is computationally expensive. In view of this we would also like to point out that 
the generation of HHG in these systems are due to
strong correlation and not mean-field in nature, therefore to describe it one requires a full diagonalization of the 
system. This restricts the system sizes that
we could access due to the exponential growth of the Hilbert space. Nevertheless, this is the state of the art calculation
and there also exist further works using exact diagonalization~\cite{jaksch_98,Zhang_2010} 
(and complementing it with Density Matrix Renormalization Group(DMRG) Ref.~\onlinecite{schollwock_05}) in bosonic model, where also convergence seems 
to be achieved within similar system sizes revealing qualitatively identical physics. All these give us
confidence that our results will be valid in the thermodynamic limit.
\bibliography{refs}{}

\end{document}